# High quality cobalt ferrite ultrathin films with large inversion parameter grown in epitaxy on Ag(001).

M. De Santis,[a*] A. Bailly,[a] I. Coates,[a] S. Grenier,[a] O. Heckmann,[b] K. Hricovini,[b] Y. Joly,[a] V. Langlais,[c] A. Y. Ramos,[a] M. C. Richter,[b] X. Torrelles,[d] S. Garaudée,[a] O. Geaymond,[a] O. Ulrich.[e]

[a]Université Grenoble Alpes, CNRS, Grenoble INP, Institut Néel, 38042 Grenoble, France
[b]LMPS, Université de Cergy-Pontoise, Neuville/Oise, 95031 Cergy-Pontoise, France
[c]CNRS, CEMES (Centre d'Elaboration des Matériaux et d'Etudes Structurales), B.P. 94347, 29 rue Jeanne Marvig, F-31055 Toulouse, France
[d]Institut de Ciència de Materials de Barcelona, ICMAB-CSIC, Bellaterra, 08193 Barcelona, Spain
[e]Université Grenoble Alpes, CEA, INAC/MEM, 38054 Grenoble, France

**Abstract**

Cobalt ferrite ultrathin films with inverse spinel structure are among the best candidates for spin-filtering at room temperature. We have fabricated high-quality epitaxial ultrathin $CoFe_2O_4$ layers on Ag(001) following a three-step method: an ultrathin metallic $CoFe_2$ alloy was first grown in coherent epitaxy on the substrate, and then treated twice with $O_2$, first at RT and then during annealing. The epitaxial orientation, the surface, interface and film structure were resolved combining LEED, STM, Auger and *in situ* GIXRD. A slight tetragonal distortion was observed, that should drive the easy magnetization axis in plane due to the large magneto-elastic coupling of such a material. The so-called inversion parameter, i.e. the Co fraction occupying octahedral sites in the ferrite spinel structure, is a key element for its spin-dependent electronic gap. It was obtained through *in-situ* x-ray resonant diffraction measurements collected at both the Co and Fe $K$ edges. The data analysis was performed using the FDMNES code and showed that Co ions are predominantly located at octahedral sites with an inversion parameter of 0.88±0.05. *Ex-situ* XPS gave an estimation in accordance with the values obtained through diffraction analysis.

## 1) Introduction

Cobalt ferrite is an insulating (ferri)magnetic oxide with a high Curie temperature ($T_c$=793 K) and a large saturation magnetization [Brabers]. Alongside its low cost, these properties make it attractive for a wide range of applications. The spinel crystal structure (space group *Fd-3m*) is comprised of a face-centered cubic (fcc) sublattice of oxygen anions in which, one eighth of the tetrahedral lattice holes and one half of the octahedral lattice holes are occupied by cations. This results in the general formula $AB_2O_4$, where A and B refer to the cations located in the tetrahedral and octahedral sites respectively. In the *normal* spinel structure, *A* sites are occupied by divalent cations, and *B* sites by trivalent cations. In the *inverse* spinel structure, the divalent cations occupy half of *B* sites and the trivalent cations occupy the remaining *A* and *B* sites [Proskurina]. For cobalt ferrite, the *inverse* structure is the most stable. However, in general, this inversion is not complete and a fraction of cobalt ions remain located at tetrahedral sites. The degree of inversion typically depends on the sample preparation conditions. Due to the ferromagnetic interactions between ions in octahedral sites, and antiferromagnetic interactions between ions in octahedral and tetrahedral sites, cobalt ferrite is ferrimagnetic. Various density of state (DOS) calculations have accordingly predicted that the electronic band gap at the Fermi level differs for majority and minority spins. [Ref. Fritsch, Caffrey, Szotek]. The size of the band gap, however, depends on the degree of inversion [Fritsch2].

The use of ferromagnetic or ferrimagnetic insulators in multilayered structures is an efficient way to generate highly spin-polarized currents due to the exponential relationship between tunneling probability and the spin-dependent barrier height. This so-called spin-filtering effect was first observed in EuS at low temperature [Moodera], but in the last decade, the research have focused on ferrites because of their high Curie temperatures, that create the possibility spin-filter tunneling at room temperature. A second important property of cobalt ferrite is its significant magnetostriction [Bozorth] that results in a large strain-dependent magnetocrystalline anisotropy energy in ultrathin epitaxial films. A compressive strain favors an in-plane magnetization as in the case of $CoFe_2O_4/MgAl_2O_4(001)$ [Matzen], while tensile strain induces a perpendicular magnetization, observed for $CoFe_2O_4/MgO(001)$ [Chambers, Lisfi Yanagihara]. These findings have been confirmed by theoretical calculations [Fritsch]. Therefore, the incorporation of cobalt ferrite in artificial multiferroic heterostructures may result in new phenomena that opens the way to a large range of applications. It was shown, for example, that an elastic strain-mediated magnetoelectric coupling can be used to reverse the magnetization in columnar $CoFe_2O_4$ nanostructures embedded in ferroelectric $BiFeO_3$ [Zavaliche].

To fine-tune the inversion parameter and induce epitaxial strain, a precise growth methodology is essential. The growth of transition metal oxides on fcc (001) metallic substrates is strongly influenced by the lattice misfit. For example CoO films grow (001) oriented on Ag(001) [Torelli] while in the case of CoO/Ir(100), (111) films are usually obtained [Meyer]. However, in this latter system the CoO orientation can be changed to (001) by depositing a Co buffer layer of ~ 2 monolayers (ML) thickness prior to oxidation [Gubo]. This layer is pseudomorphic and forms, after a moderate oxidation, a c(4x2)-$Co_3O_4$/Co/Ir(001) reconstructed surface, which acts as a precursor for the growth of CoO(001). A similar method for growing high quality (001) magnetite ultrathin films on Ag(001) has already been previously demonstrated [Lamirand]. Facilitated by a lattice mismatch of only 0.8 %, iron grows pseudomorphically on Ag(001). Provided that a few monolayers of Fe are initially deposited, the lattice expands during oxidation, but its relative orientation is maintained. Here, the same technique is employed to obtain high quality ultrathin cobalt ferrite films with a sharp interface, a relatively flat surface, and a large inversion parameter.

In the next section, the experimental setups and deposition methods are described together with a qualitative characterization of the surface. In section 3 the film structure is solved by grazing incidence x-ray diffraction (GIXRD). Sections 4 and 5 are devoted to the determination of the inversion parameter, by x-ray photoelectron spectroscopy (XPS) and resonant x-ray diffraction (RXD), respectively.

## 2) Setups, sample growth and characterization.

All films were prepared in a similar manner using one of two distinct experimental setups, both of which are fully equipped for sample preparation and analysis in an ultra-high vacuum (UHV) environment. The first setup (LEED/STM) is located at the Néel Institute and possesses a commercial scanning tunneling microscope (Omicron VT STM/AFM), a low-energy electron diffractometer (LEED) and an Auger electron spectrometer (AES). Samples grown at this location were then transferred to the LMPS laboratory in Cergy-Pontoise, where photoelectron spectra were measured using a Mg $K_\alpha$ X-ray emission source (1253.6 eV) and a hemispherical analyzer. The second setup (GIXRD) is installed at the French BM32 beamline (CRG-IF) of the European Synchrotron Radiation Facility (ESRF). This setup consists of a UHV chamber equipped with evaporation sources for MBE growth and with AES. The system is mounted on a Z-axis diffractometer with additional degrees of freedom for sample positioning provided by a hexapod. This setup was used for resonant and non-resonant x-ray diffraction experiments *in-situ*. The oxide layers were grown on a Ag(001) single crystal with a miscut of less than 0.1°. Prior to deposition, the substrate was cleaned by repeated cycles of $Ar^+$ ion sputtering followed by annealing at approximately 850 K. Cleanliness was checked by AES, such that all contaminants were below the detection limit. Iron and cobalt were evaporated from pure rods using water-cooled electron-beam evaporators. The base pressure was in the low $10^{-11}$ ($10^{-10}$) mbar range for the STM (GIXRD) setup. The Fe (Co) deposition rate was typically 1 ML per 5 minutes (10 minutes), calibrated with a quartz crystal microbalance. STM images were obtained in constant current mode using a voltage bias ($V_{sample}$) applied to the sample. The non-resonant GIXRD measurements were performed with a photon energy of 9500 eV. Resonant measurements were carried out scanning both Co and Fe $K$ absorption edge regions. To increase the signal to noise ratio, the incidence angle was set at the critical angle for total x-ray reflection for Ag at the different energies (0.37°, 0.46° and 0.50° for 9500 eV, 7709 eV, and 7112 eV, respectively). The diffraction data were collected using a 2D detector (MAXIPIX, ESRF).

A cobalt ferrite seed layer was initially prepared in the LEED/STM setup by a three-step method. Firstly, Co (2 ML as referred to the Ag surface atomic density) and Fe (4 ML) were codeposited on the substrate kept at room temperature (RT) under UHV, forming an epitaxial metallic alloy. After deposition, the oxide layer was formed by dosing with $10^{-6}$ mbar $O_2$ for 10 minutes at RT. This step is essential to avoid intermixing with Ag. Finally, the $O_2$ partial pressure was maintained whilst the sample was annealed up to 750 K for 10 minutes via an intermediate 10 minute interval at 570 K. Following the deposition of the seed layer, the film thickness was increased by reactive codeposition of cobalt and iron (2 ML and 4 ML respectively) in the presence of molecular oxygen ($10^{-6}$ mbar) at 750 K, resulting in a cobalt ferrite film about 4 nm thick. For the GIXRD experiments, the samples were grown following the same procedure, with the exception that oxygen annealing was performed up to a higher temperature of ~870 K with more temperature intervals during the heating process. During annealing, the film structure was monitored at each interval. The higher temperature was also maintained during the reactive deposition which subsequently ensured a large average crystallite domain size in the surface plane. A value of about 30 nm was found applying the Scherrer equation to the ferrite peak width.

This allows a proper measurement of the diffraction rod intensity without the need for corrections of the detector area.

Figure 1 shows AES spectra measured in the LEED/STM set-up (continuous line, red online) and in the GIXRD one (black circles) after oxide growth. The principal peaks of O, Fe and Co are labelled. The peaks at 598 and 775 eV, arising from Fe and Co levels respectively, were used to infer the ferrite stoichiometry. The two samples have the same composition within error. We measure a signal ratio AES($Fe_{598}$)/AES($Co_{775}$) of ~ 0.75, while the corresponding cross section ratio is $\sigma(Fe_{598})/\sigma(Co_{775})$ ~ 0.46. Then the ratio Fe : Co ~1.63, i.e. the ferrite is slightly enriched in Co with respect to the desired composition. As will be discussed later, the more quantitative x-ray resonant diffraction analysis gives a ratio of ~1.86. Some Ag surface segregation is observed in the sample annealed at 870 K (peak at 356 eV). A rough estimation based on the relative cross sections gives about 0.5 equivalent ML of Ag on the surface.

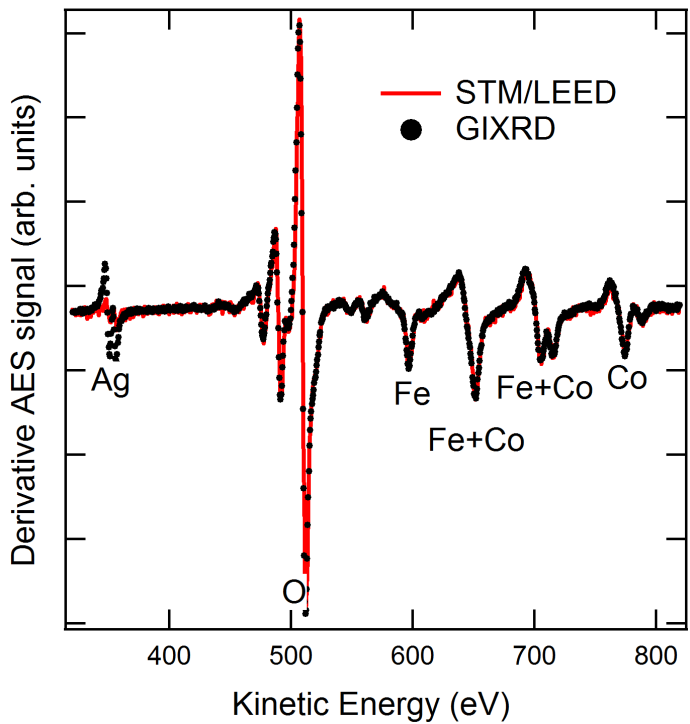

Fig.1 Derivative Auger spectra of the samples elaborated in the GIXRD chamber (black circles) and in the LEED/STM chamber (continuous line, red online).

Figure 2 shows the LEED pattern of the Ag substrate (a) and of the final oxide (b). The latter is generated by an epitaxial oxide layer with *P4mm* symmetry and lattice constant almost twice that of Ag. This fits well with a (001)-oriented $CoFe_2O_4$ film. In addition, weaker spots of two domains at 90° of a (3×1) surface reconstruction are observed.

A (30×30 $nm^2$) medium resolution STM image recorded at room temperature of the same oxide sample is shown in Fig.3. The surface is flat on such a length scale and rows spaced by about 1.7 nm, the period of the (3×1) reconstruction, are clearly observed. The step height between dark and clear regions in the figure is about 0.22+/-0.03 nm. Large size images show that this value is the most often encountered as step height (while scanning with $V_{Sample}$=2 V).

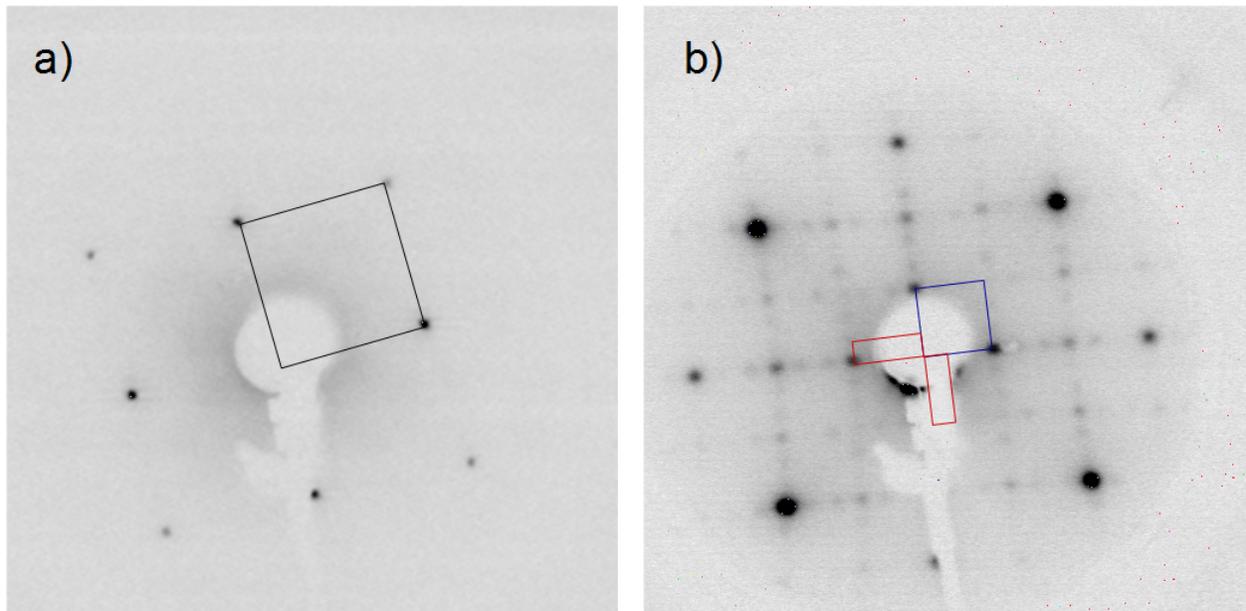

Fig. 2. LEED pattern of (a): clean Ag(001); (b): $CoFe_2O_4$/Ag(001). The ferrite films grows (001) oriented and exhibits a (3×1) surface reconstruction. The Ag, magnetite, and (3×1) reciprocal space surface unit cells are represented by the black square, blue square, and red rectangles, respectively (color version online).

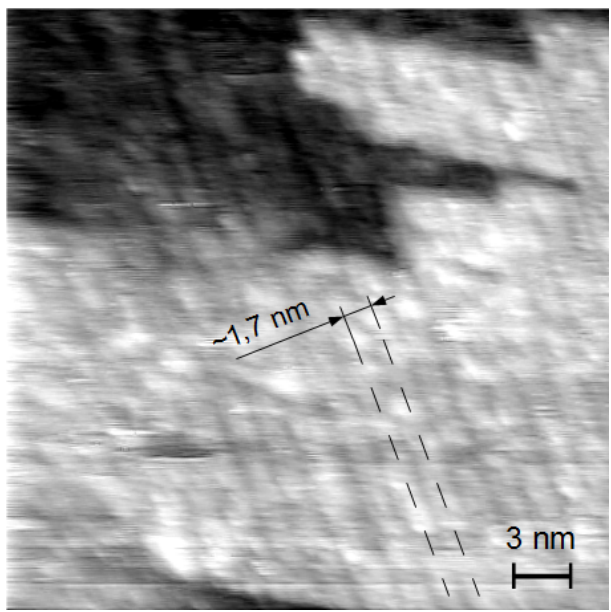

Fig. 3. STM image recorded at room temperature of the cobalt ferrite ultrathin film, 30×30 nm$^2$, $V_{Sample}$=2 V, $I_{Bias}$=0.1 nA

The ferrite structure can be viewed along the (001) direction as a stacking of eight unique layers that alternate between planes containing oxygen anions and *B*-site cations, and planes containing only *A*-site cations . The interplanar spacing between planes of the same type is 0.21 nm, which is directly comparable to the observed step height. It can therefore be concluded that the surface termination of the cobalt ferrite structure is dominated by one of these two types of planes. A similar result was observed at the surface of (001) bulk magnetite which exhibits *B*-type termination [Stanka].

### 3) XRD.

GIXRD data were collected to quantitatively solve the structure. The diffraction pattern has a 4-fold symmetry as already observed by LEED. The in-plane [100] direction is aligned with the [100] of the silver substrate, indicating epitaxial growth. The film is relaxed and its diffraction pattern does not interfere with that of the substrate. Two sets of sharp peaks are therefore observed while scanning the momentum transfer, $Q$, parallel to the surface. The first set is given by the so-called crystal truncation rods (CTRs) of Ag [Robinson2], the second set corresponds to the rods originating from the film. They are located at integer (*HK*) values, once indexed in the film mesh reference. Perpendicular to the surface, an intensity distribution, the so-called diffraction rod, is observed for each (*HK*) value [Robinson]. It exhibits wide peaks and thickness intensity oscillations. The in-plane film lattice constant $a_{film}$ was obtained by scanning across some diffraction rods at $Q_z = 2\pi/a_{cfo}$, where $a_{cfo}$, the bulk cobalt ferrite lattice constant, is 838.6 pm [Proskurina], [Mohamed]. The position of 6 rods (including 5 non-equivalent ones) was carefully measured. A linear regression analysis gives $a_{film}$ = 836±2 pm. With the objective to achieve a representative portion of reciprocal space, a large set of intensities along Bragg rods, $I_{HK}(L)$, was measured. The standard procedure is to set the diffractometer angles to define each (*HKL*) point in the sample's reciprocal space and then rock the sample azimuth to integrate the intensity across the rod at a given *L*-value. The structure factors $F_{HKL}$ are then extracted by applying standard correction factors [Vlieg]. A set of 416 reflections distributed along 14 rods was collected. The set averaged to 293 non-equivalent reflections according to the substrate's *P4mm* surface symmetry (10 non-equivalent rods). An agreement factor of 7% between equivalent reflections was found and used as systematic error estimation for the final experimental error calculation [Robinson]. The sampling interval $\Delta Q_z$ required to describe the rod shape decreases with respect to the inverse of the film thickness, and the required acquisition time increases accordingly. In this study, an initial $\Delta Q_z$ value of $0.1 \times 2\pi/a_{cfo}$ was used, which proved to be insufficient to describe the film thickness oscillations. Subsequently, line scans along each rod were collected (stationary- or *L-scans*, $\Delta Q_z = 0.02 \times 2\pi/a_{cfo}$) to which specific corrections were applied to extract the structure factors. These $F_{HKL}$ curves were interpolated and multiplied by a specific scaling factor to fit the data obtained from the rocking scans. The resulting rods are plotted in Fig.4. The $Q_z$ values have been renormalized using a factor calculated from the cobalt ferrite density, to take into account the x-ray refraction of the incident beam at the vacuum-film interface [FEIDENHANS'L].

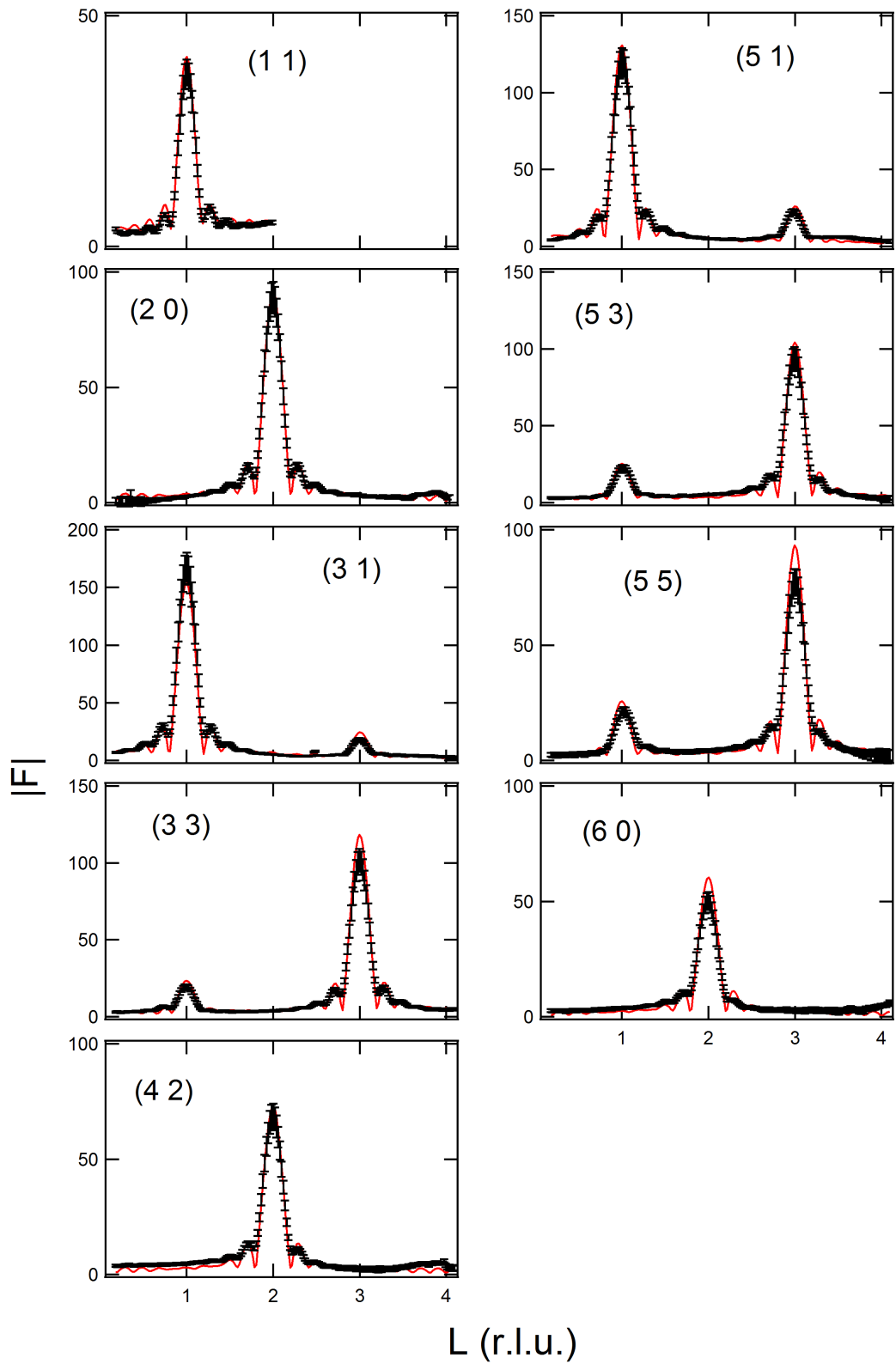

Fig. 4 Experimental CoFe$_2$O$_4$ film rods with error bars (black symbols) and best fit (continuous line, red online). The corresponding (H K) values are indicated on each panel. L is related to the $c_{film}$ lattice constant.

The first step in the analysis was the calculation of the average film interlayer spacing by fitting the peak positions along the rods. Using a linear regression based on the 14 non-equivalent peaks, $c_{film}$ value of 841±3 pm was found, resulting in a slight tetragonal distortion whereby $c_{film}/a_{film}$ = 1.006±0.006. The unit cell volume was found to be equal to that of the bulk material within the error, $(V_{film}-V_{cfo})/V_{cfo} = (-2\pm8) \times 10^{-3}$.

Before resolving the film structure, the interface roughness, that modifies the rods' shape, was studied by analyzing the Ag CTRs. A set of 47 reflections belonging to three different CTRs was collected. Their structure factors are reported in Fig.5, indexed within the Ag(001) surface cell ($\vec{a_s^1} = \frac{a_{Ag}}{2}[1\bar{1}0]$, $\vec{a_s^2} = \frac{a_{Ag}}{2}[110]$, and $\vec{a_s^3} = a_{Ag}[001]$). The error is evaluated using the agreement factor obtained from film reflections.

Since the oxide film exists in incoherent epitaxy, it does not contribute to the CTRs intensity, which depends only on the substrate parameters and the interface roughness. An initial best fit of the (11) CTR is shown in Fig. 5a (dashed line, red online). It is calculated using bulk interlayer distances and Debye-Waller (DW) factors ($B_{Ag}=0.7\times10^4$ pm$^2$). The roughness, considered within the β model [Robinson2], is therefore the only remaining physical parameter to be optimized. A value β=0.12±0.03 is obtained, resulting in a normalized $\chi^2$ of 0.84.

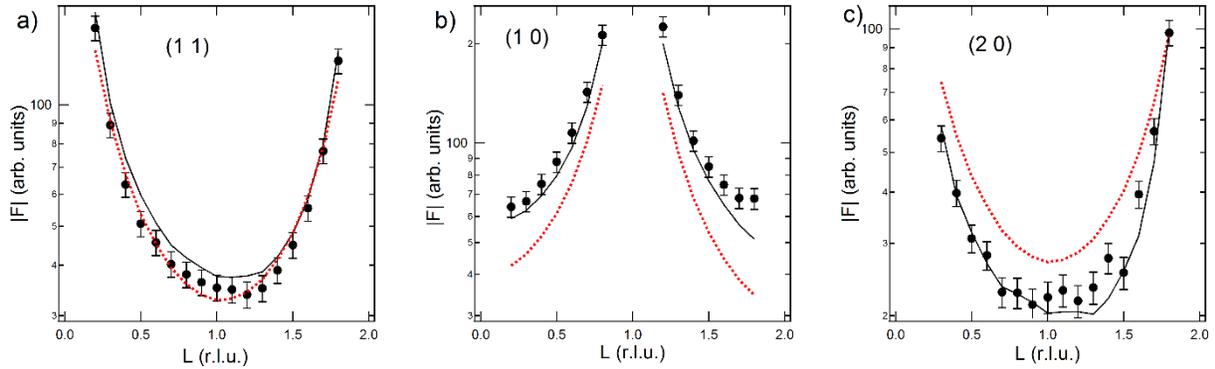

Fig. 5. Experimental Ag(001) CTRs (black symbols with error bars), calculations with best β fit only (dashed line, red online) and with best surface in-plane DW fit (continuous line).

However, as shown in Fig.5, this model fails to give an accurate fit of the relative intensity of the different rods. The $\chi^2$ increases to about 10 upon considering the full data-set of three CTRs. A better description is achieved by considering the role of the oxide film. The stress exerted by the film on the silver at the interface induces a localized displacement field in the silver substrate. Periodic displacements in the substrate give rise to satellite peaks close to the Bragg ones [Prevot]. Fig. 6 shows an *H*-scan close to the Ag (101) Bragg where such satellites are observed. The distance between satellites corresponds roughly to the period of the Moiré generated by the Ag and CoFe$_2$O$_4$ lattices.

The stress caused by the oxide film induces displacements that are predominantly parallel to the interface. Their amplitude decreases while going deeper into the crystal. A detailed analysis would require the introduction of a large superstructure cell, with a consequently large number of variables. However, the average structure can be qualitatively described by a structural disorder within the Ag surface unit cell. The CTRs were therefore analyzed further by introducing in-plane DW factors. A simultaneous fit of the CTRs (Fig.5, continuous black line) results in an in-plane Debye factor gradient over the first five atomic layers of the silver substrate at the interface ($B_{i,IP}$). The out-of-plane component was kept fixed at the bulk value (Table I). Note that the DW factors observed in the surface layers decreases the scattering factor amplitude relative to the bulk values, therefore its effect on a single rod is similar to that of increasing interface roughness. Within this model, the best fit β value is close to zero. It was also observed that the first Ag interlayer distance at the interface ($d_{Ag1}$) is slightly contracted, which accounts for the observed shift in the CTR's minima. The corresponding $\chi^2$ when considering the full data set of the three CTRs is 2.5.

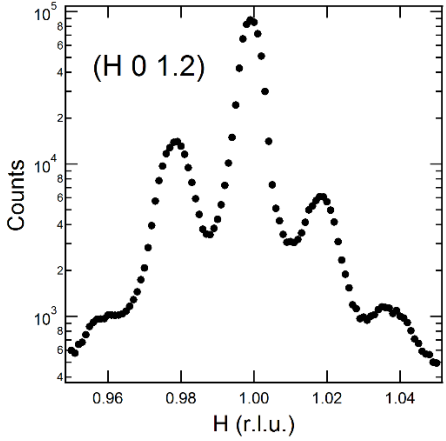

Fig.6 Ag (H 0 1.2) scan.

TABLE I Best fit structural parameters of the interfacial Ag including the interfacial interlayer distance, $d_{Ag1}$; the DW factor at the $i^{th}$ layer parallel to the surface, $B_{i,IP}$; and the roughness parameter, β.

| | |
|---|---|
| $d_{Ag1}$ (pm). | 201±1 |
| $B_{1,IP}$ (×10$^4$ pm$^2$). | 6.3±0.5 |
| $B_{2,IP}$ (×10$^4$ pm$^2$). | 3.9±0.5 |
| $B_{3,IP}$ (×10$^4$ pm$^2$). | 2.3±0.3 |
| $B_{4,IP}$ (×10$^4$ pm$^2$). | 1.4±0.3 |
| $B_{5,IP}$ (×10$^4$ pm$^2$). | 0.9±0.1 |
| β | 0.05±0.03 |
| $\chi^2$ | 2.5 |

In summary, the CTR analysis shows that the interface is quite sharp, despite the high annealing temperature, as indicated by the calculated β parameter of 0.05, which corresponds to a root mean square roughness of 43 pm [Robinson2]. In the following, the film structure is studied via the analysis of its rods considering a sharp interface.

The bulk CoFe$_2$O$_4$ unit cell is formed by eight equispaced atomic layers along the growth axis, as shown in Fig.7. The diffraction pattern of the film was calculated for a structure which follows this same layer stacking, starting with a *B* type layer at the interface (layer 1). Since the non-resonant iron and cobalt scattering factors are close in magnitude, standard diffraction methods cannot give reliable values for the film's inversion parameter, i.e. the relative Co and Fe occupancies of tetrahedral and octahedral sites. These values were obtained by XRD, as described in the following, and are used here in the best-fit of the structure. From our calculation, the film is composed of 33 complete atomic layers, on top of which, the surface is terminated by five partially occupied bilayers, corresponding to the observed surface roughness. Each bilayer was assumed to be *B* terminated, in accordance with the observed termination of the magnetite (001) surface [Stanka]. A further parameter of the crystal structure is the shift of the oxygen anion positions, $\Delta$, relative to a standard fcc sublattice in accordance with the *Fd-3m* symmetry, whereby $\Delta=\pm(0.25-u)$ times the unit cell and $u=0.261$ for the bulk material [Proskurina]. The (3×1) reconstruction rods were not observed in the GIXRD experiment. We believe that this is due, in large part, to the Ag segregation that was present in the GIXRD-system-prepared samples, which inhibits the establishment of large reconstructed surface regions. The (3×1) superstructure was therefore neglected in the model. A fixed DW factor, averaged from the Fe and Co bulk values ($B=0.5\times10^4$ pm$^2$), was used for the simulation. The most reliable film structure was obtained from the refinement agreement between experimental and calculated data using a model considering only a limited number of parameters: the bilayer occupancies ($Occ_i$), the oxygen displacement ($u$), and the first *A-B* interlayer distance at Ag interface ($d_1$). The remaining interlayer distances were kept fixed at the value $d=c_{film}/8=105.1$ pm. In the fit, the intensities of four equioccupied domains with equivalent structures rotated by 90° were added to restore the substrate symmetry. Scattering factors of Fe, Co and O ions were used. By refining these few structural parameters a relatively good agreement with the experimental data is obtained. The best-fit values are reported in Table II and the fit is plotted in Fig. 4. This fit definitively confirms the CoFe$_2$O$_4$ structure of the film. The most significant discrepancies between simulated and experimental data originate from regions close to the main peaks, where deep minima are calculated but not observed experimentally. This results in a relatively high $\chi^2$ of 6.9. It should be emphasized, however, that most of the discrepancies likely arise from the observed Ag segregation at the surface. It is known that silver deposited on the ($\sqrt{2}\times\sqrt{2}$)*R*45°- Fe$_3$O$_4$(001) surface grows both on specific crystallographic sites and forms clusters, depending on the annealing conditions [Bliem2]. A fraction of a Ag monolayer located at specific sites on the surface would correspond in modulus to a few percent of the CoFe$_2$O$_4$ film's scattering amplitude at $Q=0$. While this interference can be neglected for strong peaks, it may substantially change the diffracted intensity close to minima. Nevertheless, the comparison between the crystallographic model and the experimental x-ray diffraction data using another figure of merit, the so-called R-factor, which neglects experimental errors, gives a very reasonable value of 17%.

CoFe$_2$O$_4$ has the same spinel structure as magnetite. In the absence of 1/6 of cations in octahedral sites in magnetite, Fe$_2$O$_3$ with maghemite structure is obtained. For this reason, the octahedral site occupancy of the cobalt ferrite film was also checked. The resulting occupancy *Occ(B)* was found to be between 0.95 and 1.

Finally, TABLE III reports the experimental structure factor values obtained for each peak position together with the corresponding fitted values and the bulk $CoFe_2O_4$ values. In both cases, the agreement is fairly good.

TABLE II Structural parameters of the $CoFe_2O_4$ film

| | |
|---|---|
| $Occ_1$ | 0.85±0.05 |
| $Occ_2$ | 0.62±0.05 |
| $Occ_3$ | 0.56±0.05 |
| $Occ_4$ | 0.23±0.05 |
| $Occ_5$ | 0.23±0.05 |
| $u$ | 0.254±0.001 |
| $d_1$ (pm) | 94±1 |
| $Occ(B)$ | 1+0/-0.05 |
| $x_{A,Co}$ | 0.13±0.05 |
| $x_{B,Co}$ | 0.46±0.03 |
| $\chi^2$ | 6.9 |

TABLE III experimental, fitted and bulk $CoFe_2O_4$ structure factors.

| Reflection index | Exp. | Fit | Bulk [Proskurina] |
|---|---|---|---|
| (111) | 38±3 | 41.1 | 40.5 |
| (202) | 89±7 | 91.3 | 88.2 |
| (311) | 168±12 | 153.5 | 137.3 |
| (313) | 18±1 | 24.4 | 9.45 |
| (331) | 19±1 | 23.4 | 9.45 |
| (333) | 102±7 | 118.3 | 96.8 |
| (422) | 69±5 | 72.9 | 68.5 |
| (511) | 120±9 | 130.6 | 132.1 |
| (513) | 22±2 | 26.0 | 25.3 |
| (531) | 23±2 | 25.3 | 25.3 |
| (533) | 94±7 | 104.2 | 98.9 |
| (551) | 21±2 | 25.6 | 33.3 |
| (553) | 77±6 | 93.2 | 96.6 |
| (602) | 50±4 | 60.4 | 56.3 |
| $\chi^2$ | | 6 | 14 |

### 4) Photoemission

X-ray photoelectron spectra were collected *ex-situ* on the sample grown in the STM setup after degassing for a few hours at approximately 450 K in UHV to remove adsorbed molecules (no trace of carbon contamination was detected by XPS). Fig.8 (a) and (b) show the Co *2p* and Fe *2p* core-level photoemission regions, respectively. Each level is comprised of the $2p_{3/2}$ and $2p_{1/2}$ sublevels. Their binding energy depends on the valence state and on the local structural environment. Co ions in ferrite are divalent, and a splitting is observed for atoms in octahedral (*B*, *D3d* symmetry) and tetrahedral (*A, Td* symmetry) sites.

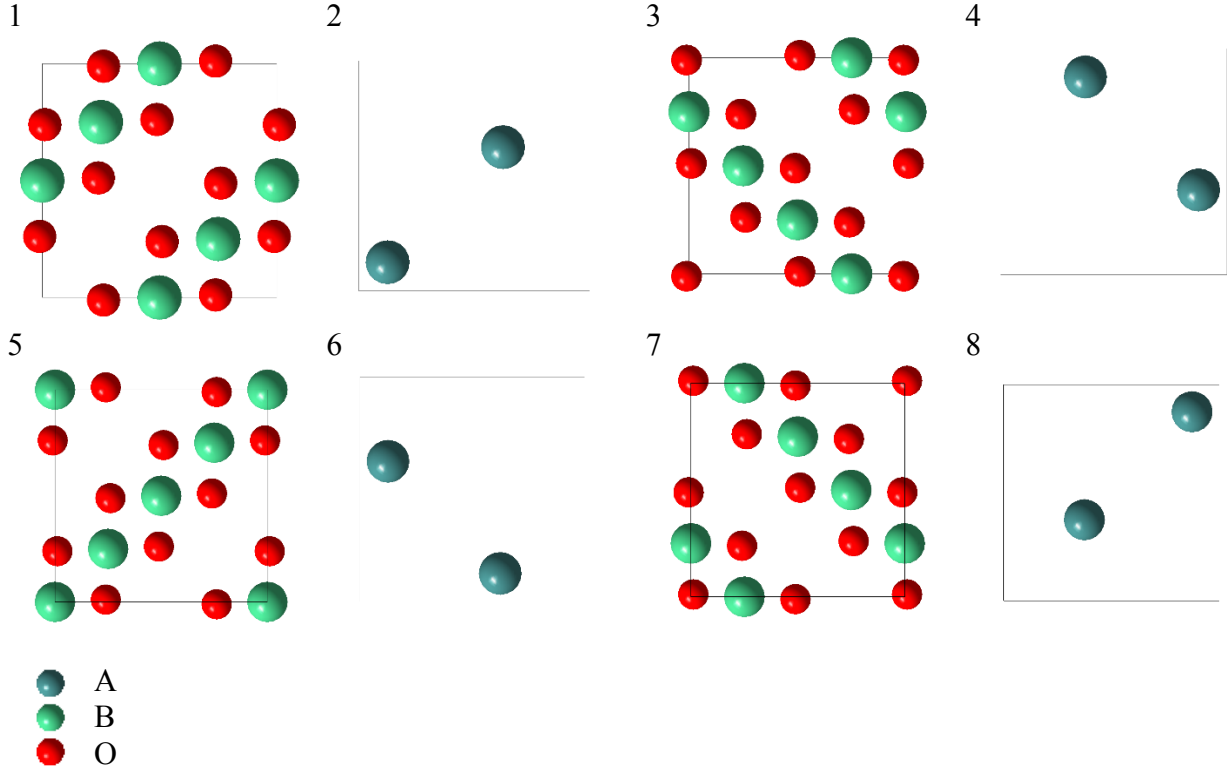

Fig. 7. Layer-by-layer structure of bulk $CoFe_2O_4$ unit cell, following the stacking order. *A* and *B* correspond to the tetrahedral and octahedral sites, respectively. The interlayer spacing is $c_{film}/8$

Each sublevel has a shake-up satellite, due to the excitation of a 3d electron by the core level photoelectron. Therefore the Co 2p level has been fitted with 6 Gaussian peaks, after subtraction of a suitable background (a Shirley background plus a linear one which represents the energy loss contribution of peaks at lower binding energy). Iron is trivalent in $CoFe_2O_4$ [Aghavnian]. The analysis of its 2*p* level is difficult due to the presence of the oxygen Auger peak and due to the fact that the shake-up satellites are poorly defined. Therefore, in this case, a reliable fit cannot be found without constraints. Aghavnian *et al.* [Aghavnian] determined the inversion degree in $CoFe_2O_4/BaTiO_3$ films by X-ray Magnetic Circular Dichroism (XMCD) and used their results to fit XPS 2*p* data. Here, for both elements, we have focused on the most reliable $2p_{3/2}$ level constraining the energy splitting between *A* and *B* sites to the values given in ref. [Aghavnian] (2.4 eV and 2.65 eV for Co and Fe, respectively). The areas of the Co $2p_{3/2}$-*D3d* and Co $2p_{3/2}$-*Td* peaks give directly the fraction of cobalt in octahedral sites, i.e. the inversion parameter. A value of about 0.79 is obtained. On the other hand, equivalent analysis of the Fe $2p_{3/2}$ doublet provides an estimation that 47% of iron is located on tetrahedral sites. These results have to be compared with those obtained by the quantitative resonant diffraction analysis (0.88±0.05 and 45±3%, respectively, see next section).

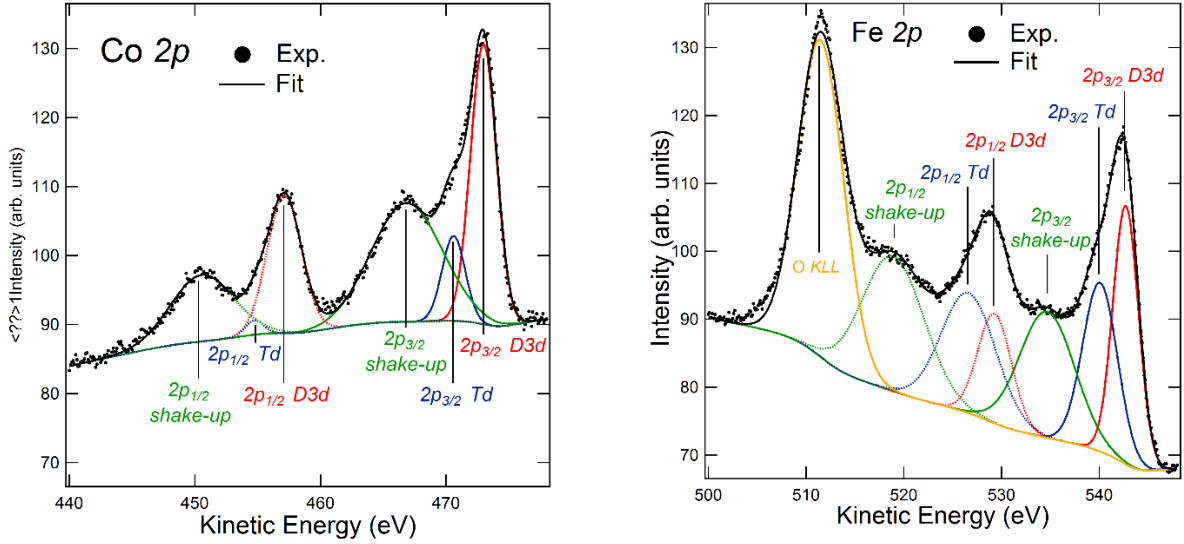

Fig. 8. Experimental Co (a) and Fe (b) 2*p* core level photoemission lines and best fits.

### 5) Resonant x-ray diffraction

RXD exploits the change in the atomic scattering factor close to an absorption edge to study the composition and the atomic environment of a given crystallographic site [Grenier]. In the present experiment, the intensity changes of several ferrite film diffraction peaks were measured by scanning the energy close to both the Fe and the Co absorption *K* edges. Experimentally, this requires the diffractometer circles to move in such a way that the (*HKL*) position is kept fixed while scanning the energy. These energy scans are shown in Fig. 9 for a set of 6 peaks at both edges and for 5 additional peaks at the iron edge only. Some of these peaks exhibit a very strong intensity variation, which makes this technique very powerful in determining the stoichiometry of the octahedral and tetrahedral sites of the spinel structure. The $CoFe_2O_4$ unit cell structure factors can be written as the sum of three contributions originating from the *A*, *B* and oxygen sites:

$$F(HKL) = F_B + F_A + F_O. \qquad (1)$$

While most of the reflections are sensitive to all atomic sites, it is easy to calculate that:

$$F(2+4n, 0, 2) = (-1)^{n+1} \times 8 f_A + F_O, \qquad (2)$$

and:

$$F(2+4n, 2, 2) = 16 f_B + F_O, \qquad (3)$$

where $f_A = f_{Co} \times x_{A,Co} + f_{Fe} \times (1-x_{A,Co})$, and $x_{A,Co}$ is the fraction of *A* sites occupied by Co (analogously for $f_B$).

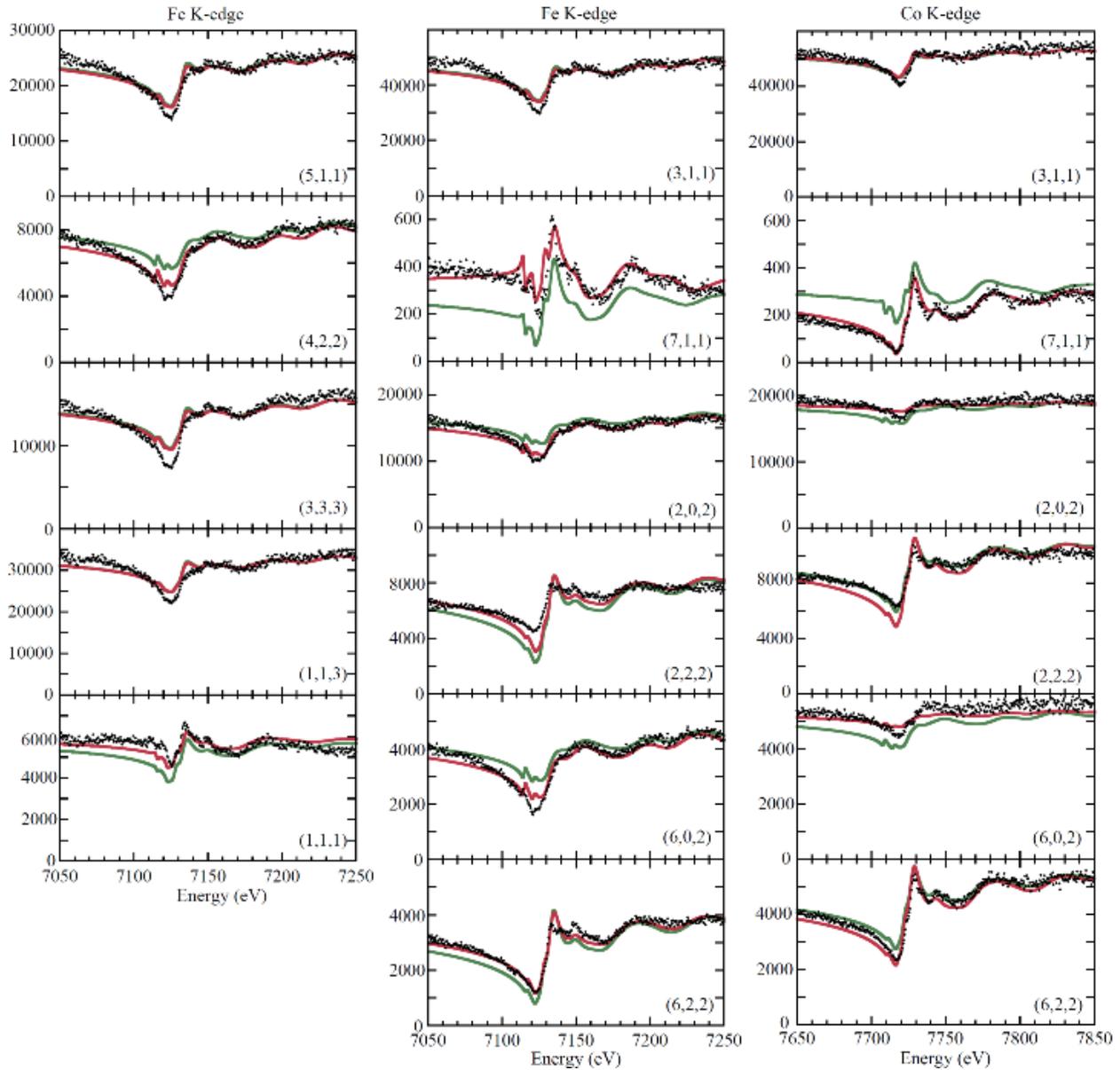

Fig.9. Experimental RXD of selected (*HKL*) reflections at the iron (left and central columns) and cobalt (right) *K* edges (black symbols), calculated intensity for best fit occupancy (continuous red lines) and for statistical occupancy (dashed green lines).

This means that the (202) and (602) reflections plotted in Fig. 9 are sensitive to the tetrahedral cation sites only, while the (222) and (622) ones to the octahedral ones only. A first inspection of the respective curves brings the conclusion that the inversion parameter is quite high, because, for example, the intensity variation of the (202) reflection is much larger at the Fe edge than at the Co one, i.e. the tetrahedral site is iron rich. A more quantitative analysis requires the precise knowledge of the resonant contribution to the scattering factors, which is very sensitive to the oxidation state and to the environment of the selected element. This resonant contribution was calculated using FDMNES [Bunau], an *ab initio* code already extensively used to simulate XANES and RXD. Its density functional theory (DFT) full potential approach makes it especially appropriate for simulating absorption edges of chemical elements embedded in non-close packed surroundings or in low symmetry sites. The specific Co and Fe scattering factors were calculated for atoms located both in octahedral and tetrahedral sites, taking a spinel structure with statistical occupancy of the cation sites. Then they were inserted in the unit cell structure factor to calculate the intensities. The occupancy of each site was obtained through a best fit of the experimental data (continuous line in Fig.9, red online). It was found that tetrahedral and octahedral sites are occupied at 13±5% and 46±3% by Co, respectively. This results in an inversion parameter of 0.88±0.05 and in an average stoichiometry of $Co_{1.05}Fe_{1.95}O_4$. In Fig.9, the intensity calculated assuming statistical occupancy of the cation sites is also shown for comparison (dashed line, green online).

## 6) Summary and conclusions.

Ultrathin cobalt ferrite layers were grown on Ag(001) by MBE following a three-step method. The films are (001) oriented and have a sharp interface, a relatively flat surface, and a bulk-like crystallographic structure. The substrate induces a slight compressive strain and the inversion parameter, determined by RXD, is close to 1. These characteristics make such films an ideal insulating barrier for spintronic applications, particularly for RT spin filtering, since the inversion parameter is linked to the height of the spin-dependent energy gap at the Fermi level [Szotek]. Theoretical calculations have also shown that strain and degree of inversion are correlated; a decrease in the lattice constant favors the inverse configuration with respect to the normal state [Fritsch2]. Therefore, the nature of the substrate and the growth mode can have a strong influence on the cation distribution. Both calculations [Fritsch] and experiments [Gao] agree that a compressive strain in cobalt ferrite films induces an in-plane magnetization. Such layers can then be combined with other transition metal oxides to form model multilayered structures whose properties can be tuned and calculated theoretically. For example, cobalt ferrite, magnetite, MgO, and other transition metal oxides can be layered-up to form magnetic tunnel junctions [Chapline] and other functional devices in coherent epitaxy. This will help to develop an understanding of the influence of the structural parameters on the physical properties.

**Acknowledgments**

Financial support through ANR EQUIPEX ANR-11-EQPX-0010 and beam time at the French CRG BM32 beamline of the ESRF are acknowledged. V. L. and X. T. also acknowledge European financial support through the EFA194/16 TNSI (POCTEFA/UE-FEDER) project.

*Corresponding author. E-mail address: Maurizio.De-Santis@neel.cnrs.fr